\documentclass{aastex}

\slugcomment{Submitted to Ap. J. Letters}
\shorttitle{X-ray Spectrum of Mrk 478}
\shortauthors{Marshall et al.}

\begin{document}

\title{The Remarkably Featureless
High Resolution X-ray Spectrum of Mrk 478}

\author{Herman L. Marshall\altaffilmark{1},
Rick A. Edelson\altaffilmark{2,3},
Simon Vaughan\altaffilmark{3,4},
Mathew Malkan\altaffilmark{2},
Paul O'Brien\altaffilmark{3},
Robert Warwick\altaffilmark{3}}
\altaffiltext{1}{Center for Space Research, Massachusetts Institute of
	Technology, 77 Massachusetts Ave., Cambridge, MA 02139}
\altaffiltext{2}{UCLA Astronomy Department, Los Angeles, CA 90095-1562}
\altaffiltext{3}{X-Ray Astronomy Group, University of Leicester, Leicester
LE1 7RH, U. K.}
\altaffiltext{4}{Institute of Astronomy, Madingly Road, Cambridge CB3 0HA, U. K.}
\email{hermanm@space.mit.edu,
rae@astro.ucla.edu, sav@ast.cam.ac.uk, malkan@astro.ucla.edu,
pto@star.le.ac.uk, rsw@star.le.ac.uk}
\slugcomment{To appear in the Astronomical Journal}

\begin{abstract}

An observation of Mrk 478 using the {\it Chandra} Low Energy Transmission
Grating Spectrometer is presented.
The source exhibited 30-40\% flux variations on timescales
of $\sim 10^{4}$ s together with a slow decline in the spectral softness
over the full 80 ks observation.
The 0.15--3.0~keV spectrum is well fitted by a single power law with
photon index $ \Gamma = 2.91 \pm 0.03 $.
Combined with high
energy data from {\it BeppoSAX}, the spectrum from 0.15 to 10 keV
is well fit as the sum of two power laws with $\Gamma = 3.03 \pm 0.04$,
which dominates below 2 keV and 1.4 $\pm$ 0.2, which dominates
above 2 keV (quoting 90\% confidence uncertainties).
No significant emission or absorption features are detected
in the high resolution spectrum,
supporting our previous findings using the Extreme Ultraviolet
Explorer but contradicting the claims of
emission lines by Hwang \& Bowyer (1997).
There is no evidence of a warm absorber, as found in the high
resolution spectra of many
Sy 1 galaxies including others classified as narrow line Sy 1
galaxies such as Mrk 478.
We suggest that the X-ray continuum may result from Comptonization
of disk thermal emission in a hot corona through a range
of optical depths.

\end{abstract}


\section{Introduction}

High resolution spectra of active galaxies have proven
to be a valuable probe of the physical processes governing the
gas near their nuclei.  The {\em Chandra} spectra of NGC 4151
\citep{ogle00} and Mrk 3 \citep{sako00},
sources in which our direct line of sight to the nucleus is subject
to heavy absorption, are characterized by strong emission lines
from hot photoionized plasmas located in the extended narrow-line
regions of these galaxies.
For an overview of the {\em Chandra} X-ray Observatory and its
complement of instruments, see \citep{weisskopf02}.
In the spectra of the broad
line Sy 1 galaxies (BLS1s) NGC 5548 \cite{kaastra00}, MCG-6-30-15
\citep{lee01}, and NGC 3783 \citep{kaspi00}, the more directly
observed power law continuum may be absorbed by highly ionized
gas that is probably associated with the broad line region.

Narrow line Seyfert 1 galaxies (NLS1s) differ
from these BLS1s in that they have narrower optical emission lines
and steeper soft X-ray continua, which are usually highly variable.
See \citet{op85} for observational criteria that define
NLS1s.
\cite{pounds95} and others have suggested that the
the accretion rates in NLS1s are closer to the Eddington limit
than in BLS1s of the same luminosities.
The high resolution {\em Chandra} grating spectra of the
NLS1 galaxies, Ton S 180 \citep{turner00} and NGC 4051
\citep{collinge01}, do not yet provide a consistent picture
of NLS1s.  NGC 4051 has a few emission
and absorption lines on a complex continuum,
while the X-ray spectrum of Ton S 180
is relatively featureless.
{\em XMM-Newton} Reflection Grating Spectrometer \citep[RGS]{rgs}
observations of NLS1s have shown broad absorption troughs
attributed to L shell transitions of weakly ionized Fe
in two cases: IRAS 13349+2438 \citep{sako01} and
Mrk 359 \citep{obrien01a}.
Further high resolution
spectra of NLS1s are needed to clarify the relation
between this class of source and their broad-line counterparts.

Mrk 478 is a classic example of the NLS1 class with a strong
steep X-ray spectrum \citep{gon94} and low interstellar column
density ($N_H$).  These attributes combine to make
it one of the brightest NLS1s in the extreme ultraviolet
(EUV, $E \sim 0.1$ keV) \citep{mfc95}.
The spectrum measured by {\it EUVE} remains
somewhat controversial: \citet{marshall96} modelled it with
a steep power law spectrum with $N_H \sim 10^{20}$ cm$^{-2}$
but \citet{hb97} later claimed that it showed strong emission lines
at 79 and 83\AA\ in the rest frame.  These differences may
have resulted from the poor statistical significance of
the spectral data.  \citet{marshall96}
argued that the rapid, high amplitude EUV variability of Mrk 478, coupled
with weak UV variability indicated that the EUV to soft X-ray
portion of the spectrum was the result of Comptonized thermal
emission near the inner edge of an accretion disk.  The {\it EUVE}
data, however, were insufficient to test for spectral variability
or to search for features in the variability spectrum on time
scales of less than a few hours.

\section{Observations and Data Reduction}

Mrk 478 was observed with the Chandra Low Energy Grating
Spectrometer (LETGS) using the High Resolution Camera
on 8-9 August 2000
(TJD = JD - 2451000. = 765.31-766.27).  The exposure time was 80857 s.
Nearly simultaneous observations were obtained with BeppoSAX
Medium Energy Concentrator Spectrometer (MECS) \citep{boella97}
from TJD 766.07 through 767.59.

\subsection{LETGS Spectral and Broad-band Variability}

\begin{figure*} 
\begin{center}
\plotone{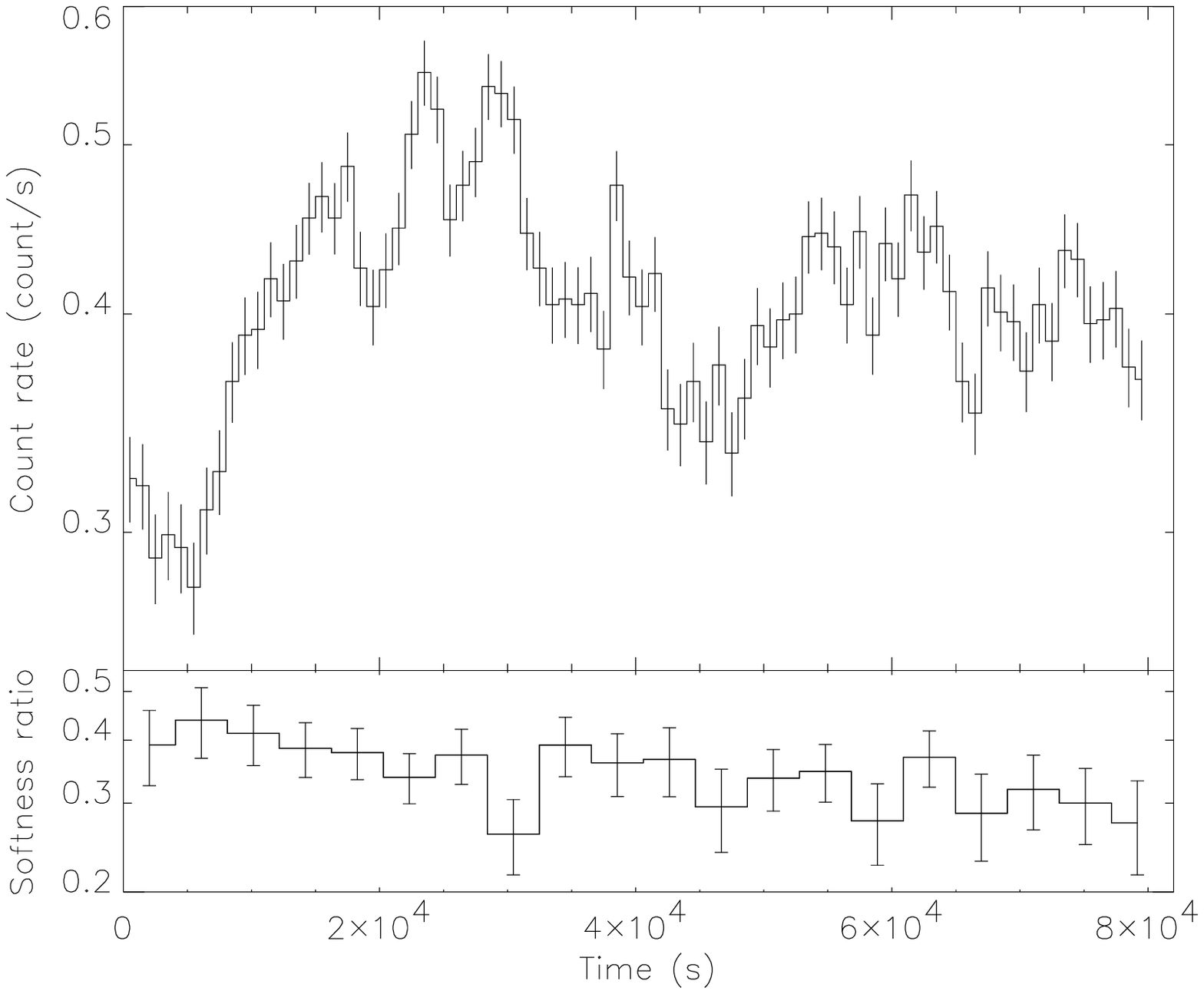}
\caption{{\em Top:} The light curve of Mrk 478 for the
LETGS observation in 300 s intervals
derived from the zeroth order data; {\em bottom:} the softness
ratio in 10000 s time intervals using the dispersed
events.
The effective bandpass of the light curve is 0.2-2.5 keV.
The softness ratio is defined as $R = (S-H)/(S+H)$, where
$H$ is the count rate in the 5-25 \AA\ (0.5-2.5 keV) band
and $S$ is defined as the count rate in
the 25-60 \AA\ (0.2-0.5 keV) band.
There is a gradual decrease in the softness
ratio as the total count rate varies sporadically on a much
shorter time scale.
\label{fig:lc} }
\end{center}
\end{figure*}

The average, background-subtracted count rate in zeroth
order was 0.409 $\pm$ 0.002 count s$^{-1}$ in a
11\arcsec\ radius aperture which contains $>$ 99\% of
the power.  Background was determined from a similarly
sized region 26\arcsec\ away from zeroth order in a
direction avoiding the dispersed spectra and the diffraction
spikes caused by the LETG coarse and fine support structure.
The count rate in 300 s bins varied by
30-40\% over the observation (see Fig.~\ref{fig:lc}).
The dispersed events were binned in 4000 s intervals in order
to determine a softness ratio, $R \equiv (S-H)/(S+H)$,
where $H$ and $S$ are the counts in hard (5-25 \AA,
or about 0.5-2.5 keV) and soft (25-60 \AA, 0.2-0.5 keV) bands.
The uncertainties were determined from counting statistics only.

The total count rate light curve showed variability on
short time scales.
Assuming that the variations follow a Gaussian distribution
with standard deviation $s$ about the mean count rate $R$, then the
fractional variability amplitude is $F_{var} = s/R = 0.132 \pm$ 0.008 after
accounting for Poisson fluctuations \citep{edelson02}.
The strongest event was an increase of almost a factor of 2 in
just 20 ksec. The softness ratio also showed a clear trend, decreasing by
about 25\% throughout the observation.  However, this slow trend is not
obviously correlated with any of the rapid variations in the total count
rate.  Fitting spectra to the first and last 10000 s of the observation
gives photon indices that changed from 3.02 $\pm$ 0.05 to
2.82 $\pm$ 0.05 over the course of the observation
(fixing the column density to the best-fit value).

\subsection{LETGS Spectroscopic Data Reduction}
\label{sec:spectra}

The LETGS spectral data were reduced from standard L1 event lists using
IDL and custom processing scripts.  The procedure is quite similar to
standard processing using CIAO and is described in more detail
by \citet{marshall01}.  The position of zeroth order was used to
locate the standard ``bow-tie'' extraction region.
The effective area (EA) of the LETGS has undergone a few revisions
since launch both due to a recalculation of the LETG efficiencies
and due to a series of in-flight calibration observations designed
to probe the HRC-S quantum efficiency.  We adjusted the EA
as described in the appendix, which provides a good fit to the
spectrum of PKS 2155-304 to a power law model.

Based on a simple model
of the spectrum (see the next section), high order fluxes are negligible
for $E >$ 0.15 keV.  In the 0.10-0.15 keV range where high orders
are important, the high order spectra are generally smooth so that
it is still possible to detect emission or absorption lines.
Based on the analysis in the appendix, the
systematic uncertainties in the high order contributions are
estimated to be $\sim$ 10\% for $E > 0.13$ keV, where
high orders contribute less than half of the observed flux.

\subsection{Continuum Fits}

\begin{figure*}
\begin{center}
\plotone{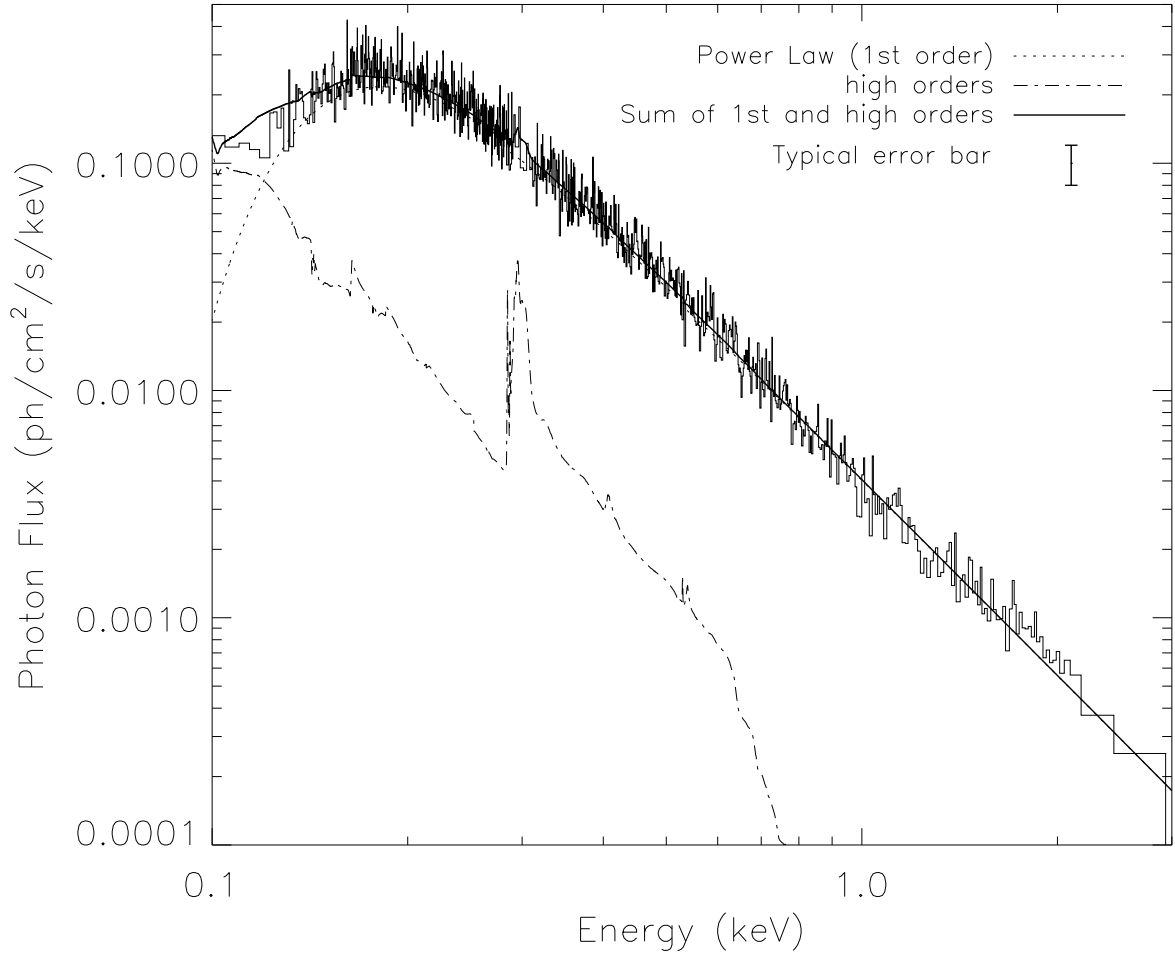}
\caption{The LETGS spectrum of Mrk 478.  The bin sizes have
been varied to provide a signal/noise ratio of
5 in each bin.
{\em Dotted line:} a power law model of the continuum
contribution from first order only, with
absorption by cold interstellar material.
{\em Dashed-dotted line:} model of the contribution
from orders 2-5, which are negligible for $E > 0.15$ keV.
{\em Solid line:} total of first and high orders, which
is an excellent fit to the data.
\label{fig:spectrum} }
\end{center}
\end{figure*}

The LETGS data were rebinned adaptively to provide a signal/noise
ratio of 5 in each bin over the 0.15 to 10.0 keV range
(see Fig.~\ref{fig:spectrum}).
Using the adjusted EA model, the spectrum was well
fitted by a simple power
law with a photon index of 2.91 $\pm$ 0.03.
Fit parameters are summarized in table~\ref{tab-spectra}.
The reduced $\chi^2$ was 1.022
for 891 degrees of freedom, which is an acceptable fit even
without adding possible systematic errors.
The quoted
uncertainties are 90\% for one interesting parameter;
the uncertainties in $N_H$ and $\Gamma$ were highly correlated
so the joint uncertainties are somewhat larger.  This model
was used as the basis for line searches in the LETGS data.

\begin{figure}
\epsscale{0.8}
\plotone{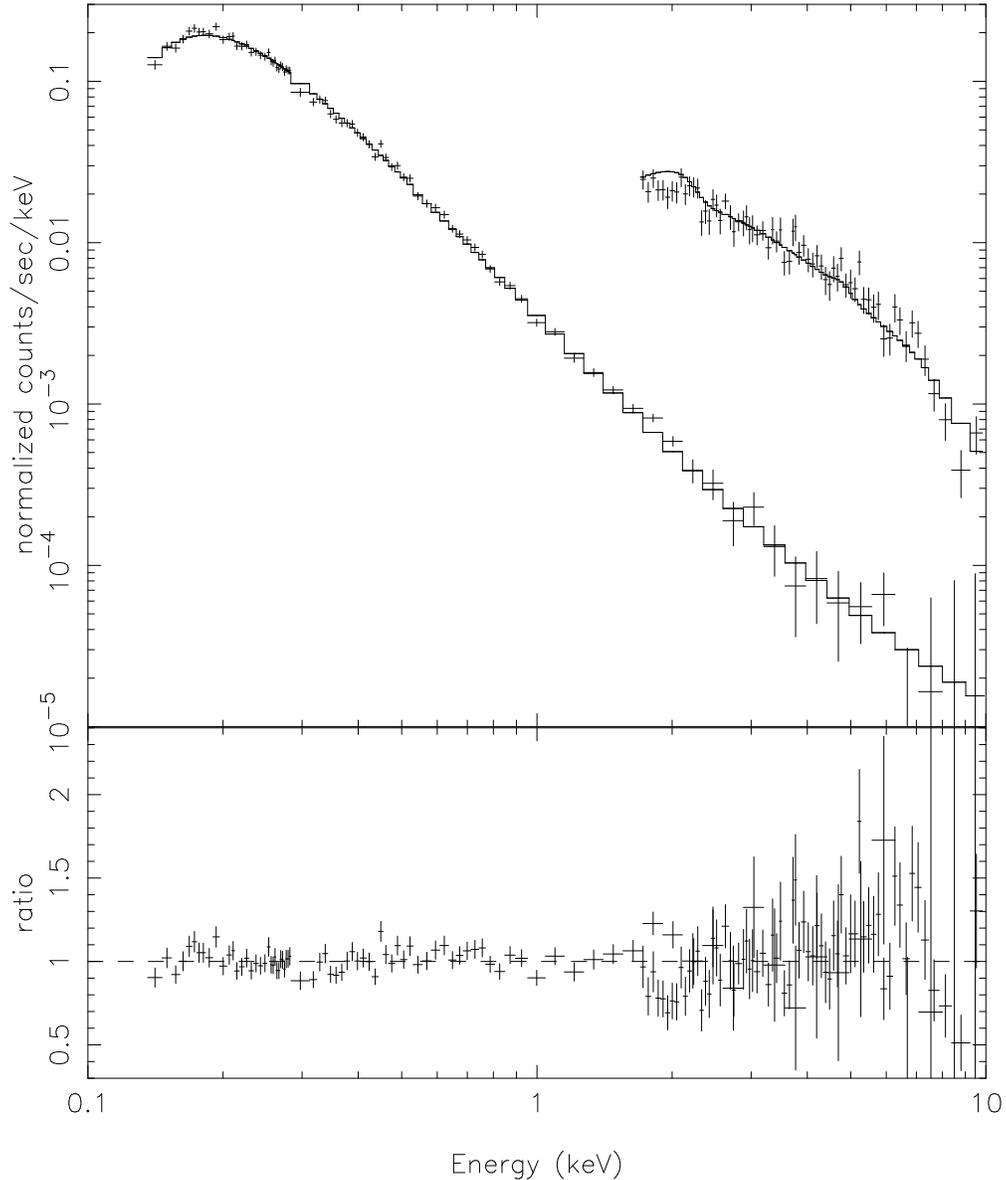}
\caption{The LETGS (left) and MECS (right) spectra
of Mrk 478 ({\em top panel})
and the residuals from the best fit model ({\em bottom panel}).
The LETGS data (spanning 0.15 to 10 keV) were rebinned
to provide a signal/noise ratio of
20 in each bin or a maximum binwidth of 5\%
and were corrected by the estimated
contributions of high orders (see Fig.~\ref{fig:spectrum}).
The model consists of the sum of two power laws with
absorption by cold interstellar material
(see table~\ref{tab-spectra}).
\label{fig:jointspectrum} }
\end{figure}

\begin{figure}
\epsscale{1.0}
\plotone{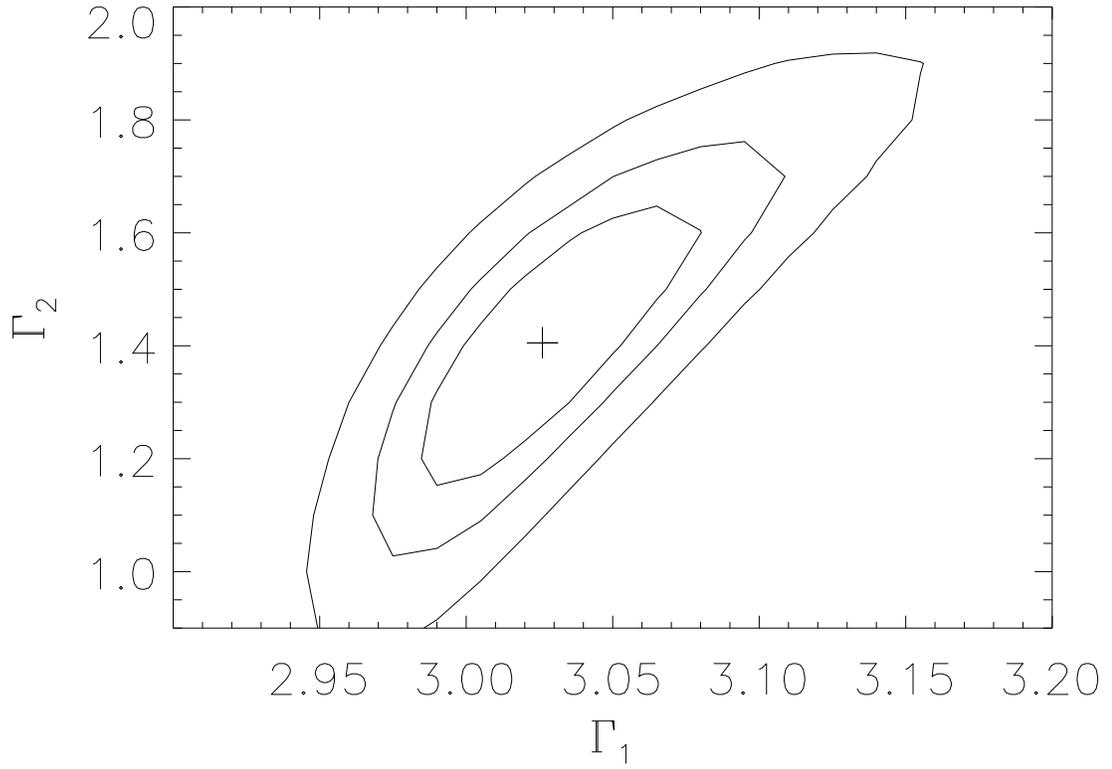}
\caption{The confidence region for the two
spectral slopes fitted to the LETGS and MECS data.
The contours give $1\sigma$, 90\% confidence, and
99\% confidence for the two parameters jointly.
\label{fig:confregion} }
\end{figure}

A single power law model was a bad fit to the LETGS and MECS data
jointly.  Combining a more coarsely binned LETGS spectrum (to
a signal/noise ratio of 20)
with the MECS data, the minimum $\chi^2$ was 439 for 152
degrees of freedom for model consisting of a single power law
(see table~\ref{tab-spectra}).  The
main deviations are found at high energies as the spectrum
flattens in the 2-10 keV band; $\Gamma = 1.98 \pm$ 0.03 was
determined from {\em ASCA} observations \citep{rt00}.
For the MECS data alone,
we obtain $\Gamma = 2.2 \pm$ 0.2, comparable to the {\em ASCA}
results and significantly flatter than the slope obtained
from the LETGS data alone.
The X-ray spectrum from 0.15-10.0 keV is best fit by the sum
of two power laws with photon indices of 3.03 $\pm$ 0.04
and 1.4 $\pm$ 0.2 (see Fig.~\ref{fig:jointspectrum}).
The joint confidence region for the two power law indices
is shown in Fig.~\ref{fig:confregion}.
The steep spectral slope of the soft
band is comparable to previous results.  Using {\em EUVE}
data in the wavelength range 65-120\AA\ (about 0.1-0.2 keV),
\cite{marshall96} obtained a slope of 4.2 $\pm$ 0.7.
\citep{gon94} also found a steep
photon index (3.32$_{-0.13}^{+0.25}$) using
ROSAT PSPC data from the 0.11-1.9 keV band.
The Galactic $N_H$ measured by 21 cm absorption \citep{mlle96} is
$9.6 \pm 1.0 \times 10^{19}$, which is consistent
with the fit results.

\subsection{Emission and Absorption Line Searches}

\begin{figure*}
\plotone{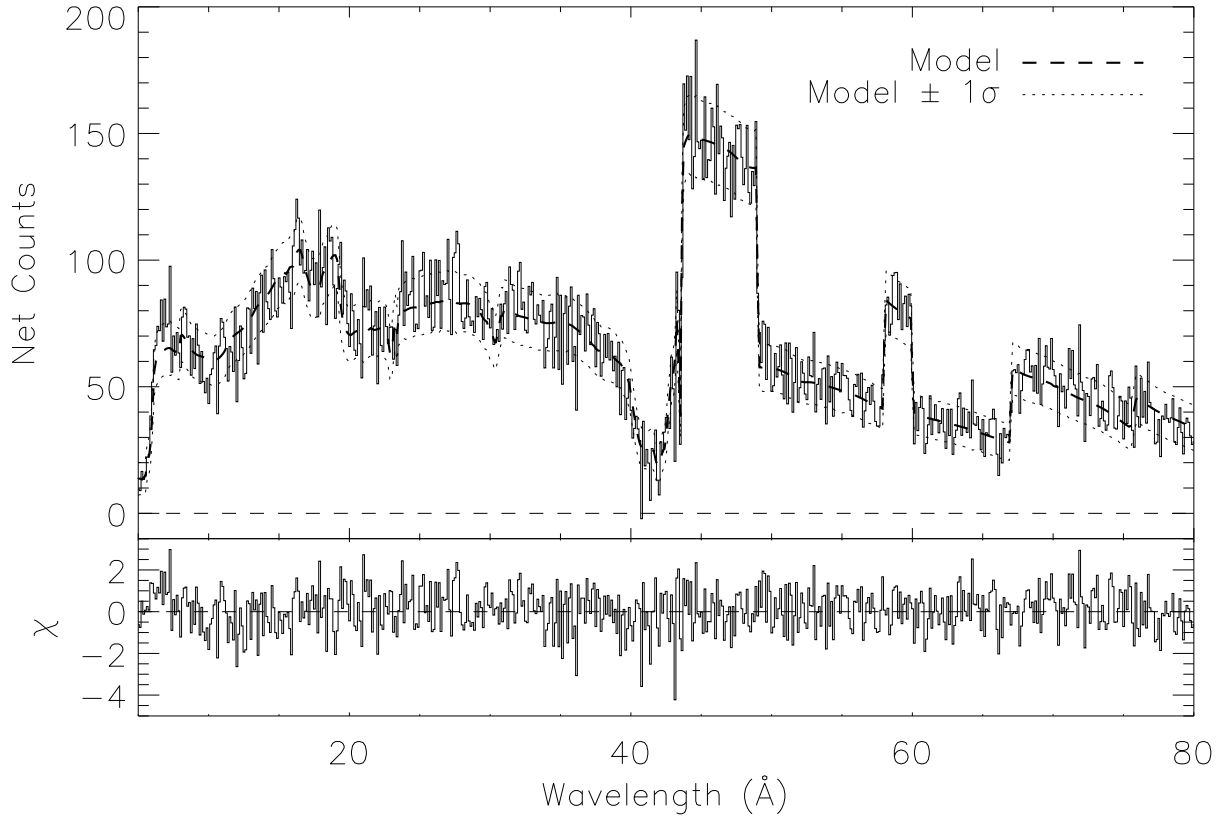}
\caption{ {\em Top:} The count spectrum of Mrk 478 obtained with the
LETGS.  {\em Bottom:} Residuals from the best fit power law
model.  A binning of 0.125\AA\ was used to obtain sufficient
signal per bin to search for narrow features.
{\em Heavy dashed line:} expected count spectrum from a steep
power law fitted to the data shown in figure~\ref{fig:spectrum}.
{\em Light dotted lines:} $\pm$ 1 $\sigma$ uncertainties about
the model.  The residuals are generally consistent with statistical
fluctuations about the model.  The sharp edges in
the model near 50 and 60\AA\ are due to detector gaps.
The only feature with sufficient significance ($> 4 \sigma$)
is a result of poor modelling of the instrumental C-K edge near
43.2 \AA.
\label{fig:countspec} }
\end{figure*}

The LETGS spectrum was binned at 0.025 \AA, which
is half of the instrument resolution.
No significant features were found
that were consistent with the instrument line response.
The data were rebinned at 0.125\AA\ to create a
count spectrum at slightly coarser resolution (Fig.~\ref{fig:countspec})
with sufficient counts per bin to
search for narrow spectral features against the continuum model.
In order to allow $<5$\% chance of a random line detection,
features must be more significant than $4.0 \sigma$, given that
about 790 bins are examined (in the 1-100 \AA\ band shown in
Fig.~\ref{fig:countspec}.  The one bin exceeding this threshold is
in the C-K edge where the EA has a systematic error.  We conclude
that there are no significant, narrow absorption or
emission features in the spectrum.  At specific wavelengths,
we can reduce the significance level threshold
because there aren't as many bins tested.
We derived 3$\sigma$ limits to the equivalent widths of unresolved
lines (FWHM $\le$ 0.05 \AA): 0.20 \AA\ (800 eV) at Fe-K$\alpha$,
0.10 \AA\ for the 4-90 \AA\ range and 0.035 \AA\ in the
6-35 \AA\ range.

\begin{figure*}
\epsscale{0.7}
\plotone{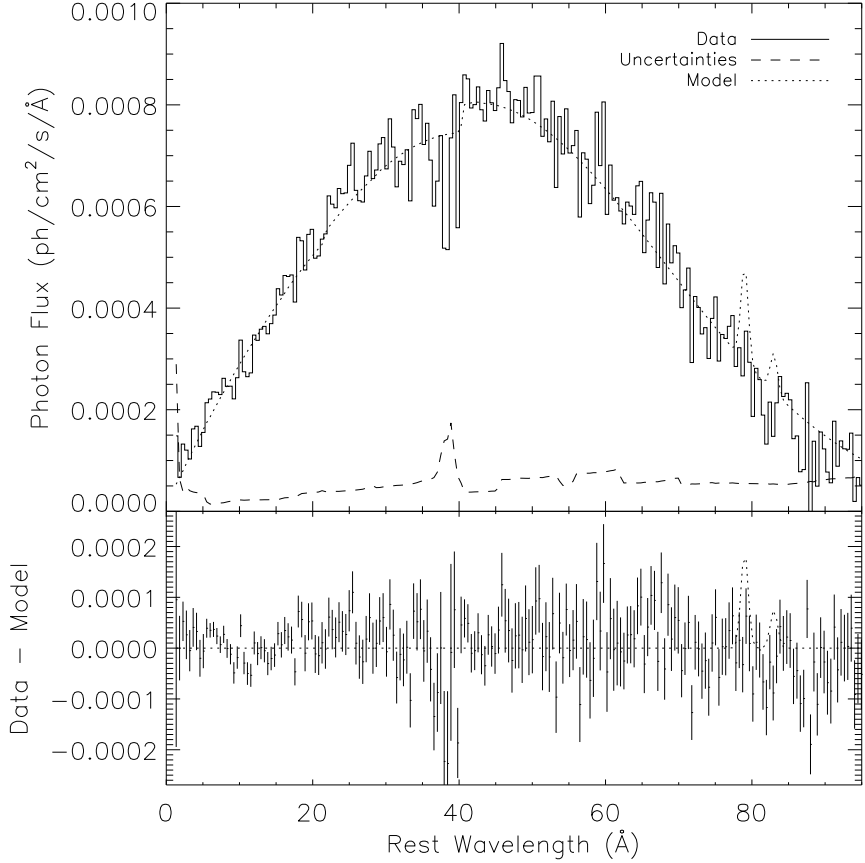}
\caption{The first order spectrum of Mrk 478 binned at
0.5\AA\ (top panel) and the residuals from the best fit power law model
(bottom panel), excluding the lines.
The contributions from orders 2-5 have been
subtracted using the power law spectral model and the
efficiencies derived from the calibration observation
of PKS 2155-304.
Residuals near the C K edge in first
order result from systematic errors in the effective
area calibration.
No broad lines are detected.
By way of illustration, the lines claimed by \citet{hb97} on the
basis of EUVE data are included in the model in the top panel (but not used
to calculate the residuals in the bottom panel). It is also clear that the
75-100 \AA\ continuum is not dominated by emission lines, contrary to the
claim of \citet{hb97}.
\label{fig:euvspec} }
\end{figure*}

Finally, we searched for broader lines, particularly in the
EUV region where \citet{hb97} reported 1-2\AA\ wide
lines at 79 and 83 \AA\ in the rest frame with fluxes
of 4.8 and 2.5 $\times 10^{-4}$ ph cm$^{-2}$ s$^{-1}$.
The line detections were claimed to be significant at the 2.8 and 3.6
$\sigma$ level, respectively.
In order to search for these features in the
Chandra spectrum, the putative lines were modeled as Gaussians
with $\sigma = 0.5$ \AA.
Using the best fit power law model and the high order efficiencies,
we computed the high order contributions and subtracted them to
produce a spectrum representative of first order only.  This
spectrum was binned to 0.5 \AA\ resolution in order to search
for lines that might be $\sim$ 1 \AA\ wide.  Fig.~\ref{fig:euvspec}
shows that any lines with $\lambda > 60 $\AA\ must be rather weak.
We can place conservative $3 \sigma$ upper limits to lines
1 \AA\ wide at $1.0 \times 10^{-4}$ ph cm$^{-2}$ s$^{-1}$
in the 60-100 \AA\ range.

\section{Discussion and Summary}

The LETGS data can give confidence to an extrapolation from the
EUV into the X-ray band, as indicated from simultaneous observations
with IUE, EUVE, and ASCA \cite{marshall02}.  It seems rather clear
now that the peak of the spectral energy distribution is in the EUV
band, as suggested by \citet{marshall96}.  The thermal peak of
disk emission is limited to less than 0.1 keV in order that
there be no significant spectral curvature in the LETGS spectrum
above 0.2 keV.  Thus, we may place a limit on the blackbody
temperature of $kT \la 0.03$ keV.

The power law shape in the .15-10 keV band can be obtained by
Comptonizing the disk thermal emission as suggested by
\citet{pounds95} in a disk corona.
We modelled the disk spectrum as a blackbody at
$kT_d = $20 eV and set the coronal electron temperature to
$kT_c \sim 40$ keV.
Using the
{\tt comptt} model \citep{titarchuk94} in {\em Xspec}, we estimate that
the optical depth can be of $\sim 0.5$
in order to obtain the rather steep spectral slope below 1 keV.
Above 1 keV, the flatter spectrum can be reproduced with the same
input spectrum and Comptonizing region but with a somewhat
larger optical depth, $\sim 5$.  In reality, there would likely
be a wide range of optical depths.  The relative strengths
of the two components would depend entirely on the relative
covering fractions; in this case, the higher optical depth
region should be $\sim$1\% of the covering fraction in at
lower optical depth.
Alternatively, \citet{hubeny01} found that self-Comptonized disk
atmospheres consisting of H and He can
have steep soft spectra for black hole masses $M \sim 10^8$ M$_{\sun}$.
Disk atmospheres with significant metal content would be
expected to show absorption edges that are not observed and have
somewhat flatter spectra \citep{rfm92}.
The spectrum of a Comptonized disk should
drop exponentially near 1 keV (Hubeny et al.) but this is where
a hard power law component begins to dominate, which would
result from Comptonization of the disk emission by a very hot corona.

It is interesting that the softness ratios vary
on relatively long time scales for both NLS1s and BLS1s.
In BLS1s, the spectra tend to soften as they brighten.  This
has been taken to indicate that a relatively slowly varying soft
component, possibly direct emission from an accretion disk, contributes to
the soft X-ray spectra of BLS1s.  In Mrk 478 and in the
NLS1s Akn~564 and Ton~S180 \citep{edelson02}, however, the total count rate is
rapidly variable and appears to be uncorrelated with the softness ratio.
The spectral modelling suggests that the entire X-ray spectrum
is governed by Compton scattering of thermal emission from the
accretion disk.  In the covering fraction model, the spectral
shapes are dependent primarily on the structure of the corona, which
is substantially larger than the hottest regions of the accretion
disk, so the spectral shape can vary slowly as the seed photon rate
varies.  In the disk self-Comptonization model, however, the
Comptonizing region is intimately linked to the photon source via
the vertical structure of the disk.  It would seem unlikely that
this vertical structure would vary slowly as the accretion rate
through the disk changes rapidly.

We find no narrow emission or absorption features.  In particular,
we do not detect emission lines at 79 and 83\AA\ that were reported
by \citet{hb97} using the same {\em EUVE} data used by \cite{marshall96}
to fit a simple continuum model.  
Similarly, none of the forbidden lines found in the NGC 4051 spectrum are
detected in our spectrum; the limit
on the Si {\sc xiii} line at 6.741 \AA\ is consistent with
the equivalent width of this line in NGC 4051 but the Ne {\sc ix}
line is at least
50\% weaker (with a limit of 0.025 \AA) and the O {\sc vii}
22.102 \AA\ line is $\times$6 weaker than their counterparts
in NGC 4051.  \cite{collinge01} show that the lines observed in
the NGC 4051 X-ray spectrum could result from a warm absorption
system as found in many broad line Sy 1 galaxies \citep{reynolds97}.
The NLS1 Akn 564 also shows signs of a warm absorber \citep{matsumoto01}.
For Mrk 478, however, we find no evidence of warm
absorption that would link this source to other NLS1s,
so warm absorption is not a characteristic
that can be used to distinguish broad and narrow line Sy 1 galaxies.
One NLS1 which has a spectrum similar to that of Mrk 478 is
PKS 0558-504.  The RGS spectrum \cite{obrien01b} shows no significant
absorption or emission line features in the 0.3-2 keV band.
Instead of fitting the soft spectrum to a power law shape, they
used a combination of several blackbody spectra.

\acknowledgments

Support for this work was provided by the National Aeronautics and Space
Administration (NASA) through Chandra Award Number NAG 5-10032 issued by the
Chandra X-Ray Center (CXC), which is operated by the Smithsonian
Astrophysical Observatory for and on behalf of NASA under contract NAS
8-39073.  HLM was supported under NASA contract SAO SV1-61010 for the CXC.

\appendix

\section{Assessing and Adjusting the LETGS Effective Area}

We started with the
LETGS effective area (EA) that was released by
the Chandra X-ray Center on 9 Mar 2000 
and updated 31 Oct 2000.\footnote{The effective area is available at
{\tt http://cxc.harvard.edu/cal/Links/Letg/User
/Hrc\_QE/EA/correct\_ea/letgs\_NOGAP\_EA\_001031.mod}.}
The LETG efficiencies were used, along with the transmission models
of the UV ion shield, to determine the EAs for orders 2-5.

The EAs were adjusted based on an analysis
of the LETGS spectrum of PKS~2155-304, observation
ID 331.  The spectrum of PKS~2155-304 was
well fitted by a power law with a photon index of 2.45 for energies
in the 0.5 to 3 keV range, so the model was extrapolated below 0.5 keV to
examine the accuracy of calibration near and below the C-K edge.
A very good fit with systematic errors of less than 10\% could
be obtained with slight adjustments to the 1st order EA model;
decreasing the optical depth at C-K (0.285 keV) by 0.13 and adding
an edge at 75 \AA\ (0.162 keV) with an optical depth of 0.4.
The 75 \AA\ edge is
an approximate description for an EA adjustment that has now been
added to the LETGS EA model but it is not a physical one.  This
wavelength merely corresponds to a point where the low energy
calibration based on HZ 43 was joined to a
high energy calibration based on PKS~2155-304 \citep{letgcal}.

Higher order
efficiencies in 2nd, 3rd, and 4th orders were adjusted by factors
of 0.5, 0.8, and 1.8, respectively, in order to match the C-K
edge features of 2nd and 3rd orders (at about 87 \AA\ and
130 \AA, respectively) and to match the overall count spectrum
in the 0.1-0.15 keV band (where these orders dominate).  We included
5th order, which was always less than 10\% of the observed counts
for $E > 0.1$ keV and ignored all higher orders.
There were significant residuals in the LETGS spectra of
both PKS 2155-304 and Mrk 421 (observation ID 1715)
at the C-K edge in first order,
however, due to a 0.5 eV error in the energies of
the resonance features (resulting from a flight filter
wavelength calibration inaccuracy) as described by
\citet{letgcal}.

\begin{figure*}
\epsscale{0.65}
\plotone{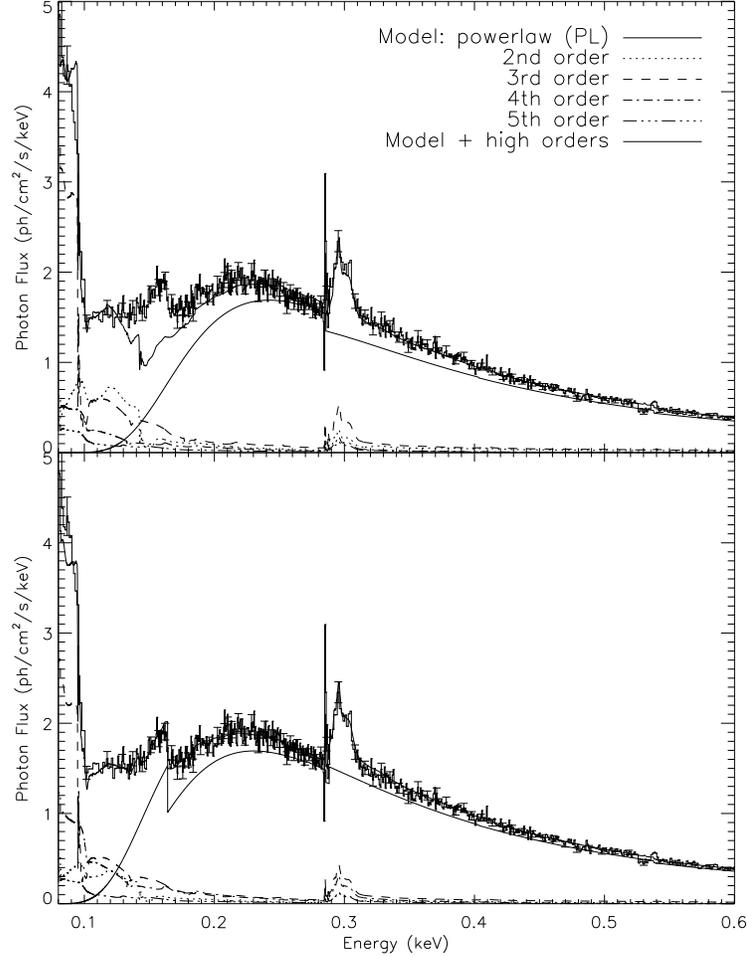}
\caption{The LETGS spectrum of PKS 2155-304 taken from
observation 331, used to assess the LETGS calibration
at low energies.  The bin sizes have
been varied to provide a signal/noise ratio of
20 in each bin.
{\em Top panel:} Spectrum computed using eq.~\ref{eq:allorder}
and the LETGS calibration files as referenced in the appendix.
High orders are computed using the fitted model and
eq.~\ref{eq:highorder}.  Several problems are very clear:
1) a continuum mismatch across the C-K edge (0.285 keV),
2) a jump at 0.162 keV,
3) the 2nd order C-K edge is too large, and
4) the 3rd order C-K edge is slightly too large.
{\em Bottom panel:} The spectrum was computed the same way
but adjustments to the high orders have been included as
described in the appendix.  Two features have been
modelled as model edges -- at 0.285 and
0.162 keV -- and then applied to the first order effective area.
The result is a very good fit to the overall spectrum from
0.08 to 0.6 keV.
\label{fig:pksspec} }
\end{figure*}

For an observation of exposure time $T$, the net
count spectrum, $C_{\lambda}$, in a $\delta\lambda$ interval
about $\lambda$ (determined for first order) is given by

\begin{equation}
C_{\lambda} = T \delta\lambda \sum_{m=0}^{\infty} m N_{\lambda/m}
	A_m(\lambda/m) ,
\label{eq:sumorder}
\end{equation}

\noindent
where $\lambda / m$ is the wavelength in order $m$
that disperses to the same physical location as $\lambda$ in
first order, $N_{\lambda}$ is the photon flux at
$\lambda$, and $A_m(\lambda)$ is the EA in order $m$ at
$\lambda$.
The spectrum shown in Fig.~\ref{fig:pksspec} was computed using

\begin{eqnarray}
n_{\lambda} & = & \frac{ C_{\lambda} } { T A_{\lambda} \delta\lambda }
\label{eq:allorder} \\
& = & N_{\lambda} + \sum_{m=2}^{\infty}
	m N_{\lambda/m} \frac{A_m(\lambda/m)}{A_{\lambda}}\;,
\end{eqnarray}

\noindent
where $A_{\lambda} = A_1{\lambda}$, so $n_{\lambda}$ is actually
an estimate of $N_{\lambda}$, differing only in the inclusion of
higher orders.
The contributions to the observed flux densities
calculated using eq.~\ref{eq:allorder}
due to order $m$ at first order wavelength $\lambda$ are

\begin{equation}
n_{\lambda,m} = m N_{\lambda/m} \frac{A_m(\lambda/m) }{ A_{\lambda} }\; .
\label{eq:highorder}
\end{equation}

\noindent
This procedure results in an ``unfolded'' spectrum in which the
computed flux density follows the true flux density closely
in the spectral region where first order dominates; i.e., where 

\begin{equation}
n_{\lambda} \gg \sum_{m=2}^{\infty} n_{\lambda,m} \; .
\end{equation}

\noindent
In spectral fitting, eq.~\ref{eq:sumorder} is solved iteratively
for $N_{\lambda}$ or the parameters of a model giving $N_{\lambda}$.
We perform all model fits this way and then use eqs.~\ref{eq:allorder}
and \ref{eq:highorder} for visualization purposes.

\clearpage

\clearpage

\begin{deluxetable}{lclclcc}
\tabletypesize{\scriptsize}
\tablecaption{Spectral Fit Parameters \label{tab-spectra} }
\tablecolumns{7}
\tablewidth{0pc}
\tablehead{
\colhead{Data Set} & \colhead{$N_H$}
	& \colhead{$K_1$} & \colhead{$\Gamma_1$} 
	& \colhead{$K_2$} & \colhead{$\Gamma_2$} & \colhead{$\chi^2$(dof)} \\
\colhead{} & \colhead{(10$^{19}$ cm$^{-2}$)}
	& \colhead{(ph cm$^{-2}$ s$^{-1}$)} &  
	\colhead{} & \colhead{(ph cm$^{-2}$ s$^{-1}$)} &
	\colhead{} & \colhead{} }
\startdata
LETGS and MECS & 9.8$\pm$0.3  &  3.32$\pm$0.22 $\times 10^{-3}$ & 3.03$\pm$0.04
	&  0.28$\pm$0.24 $\times 10^{-3}$  &   1.4$\pm$0.2  & 216(150) \\
LETGS only & 9.1$\pm$0.3  & 3.75$\pm$0.07  & 2.91$\pm$0.03  & \nodata &
	\nodata & 81.6(52)\\
MECS only  & 9.8(f) & 2.0$\pm$0.2  &  2.19$\pm$0.05 & \nodata & \nodata
	& 110.9(98) \\
LETGS and MECS & 2.6$\pm$1.4  &  3.63$\pm$0.06 & 2.79$\pm$0.02
	&  \nodata &   \nodata  & 439(152) \\
\enddata
\tablecomments{Uncertainties are 90\% confidence values for 1
interesting parameter for
a model of the form $n_{E} = e^{-N_H \sigma(E)} [ K_1 E^{-{\Gamma_1}} +
K_2 E^{-{\Gamma_2}} ] $, where $\sigma(E)$ is the energy-dependent
cross section for interstellar material of cosmic abundances.
Only one power law is fitted to the LETGS or MECS data in isolation.
An ``f'' indicates where the $N_H$ was fixed.  The number of degrees
of freedom in the fit is given in parentheses.}
\end{deluxetable}

\end{document}